\documentclass[prb,twocolumn,amsfonts,amssymb,amsmath,aps,superscriptaddress]{revtex4}

\usepackage{colordvi}
\usepackage[dvips]{color}
\usepackage{graphicx}
\usepackage{bm}

\def\hc{\mathrm{h.c.}}
\def\nn{\nonumber\\}
\def\up{\uparrow}
\def\down{\downarrow}

\def\D{\Delta}

\def\S{\Sigma}

\def\a{\alpha}
\def\b{\beta}
\def\g{\gamma}
\def\d{\delta}
\def\e{\epsilon}

\def\et{\eta}

\def\x{\xi}

\def\r{\rho}

\def\s{\sigma}

\def\bk{\mathbf{k}}
\def\ba{\mathbf{a}}
\def\br{\mathbf{r}}
\def\mpp{\mathrm{pp}}
\def\mph{\mathrm{ph}}

\def\tra{\includegraphics[width=2.5mm]{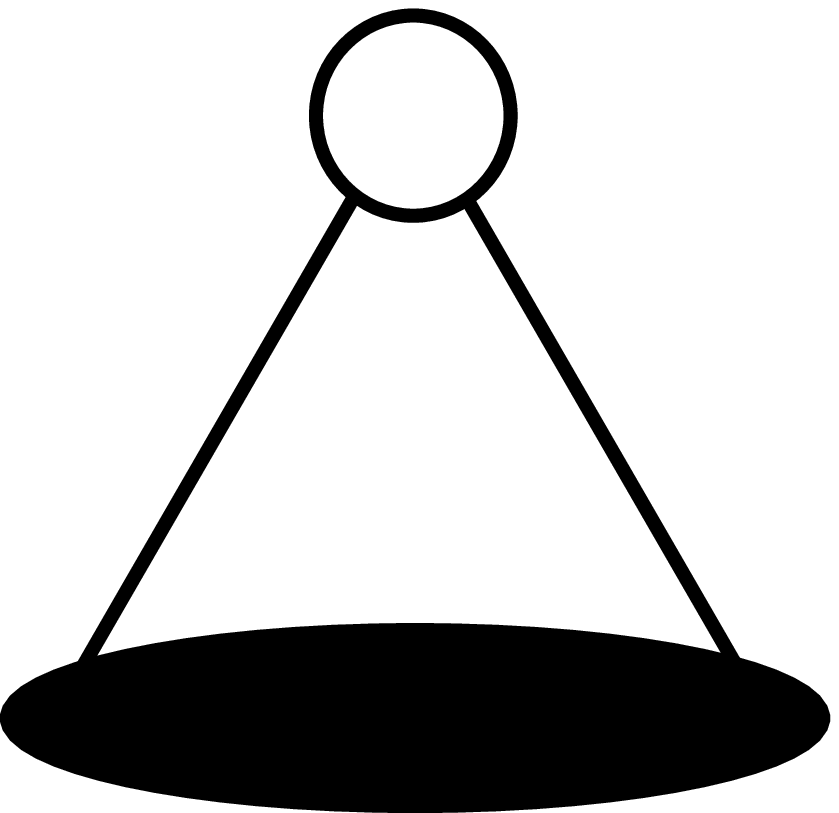}}
\def\tro{\includegraphics[width=2.5mm]{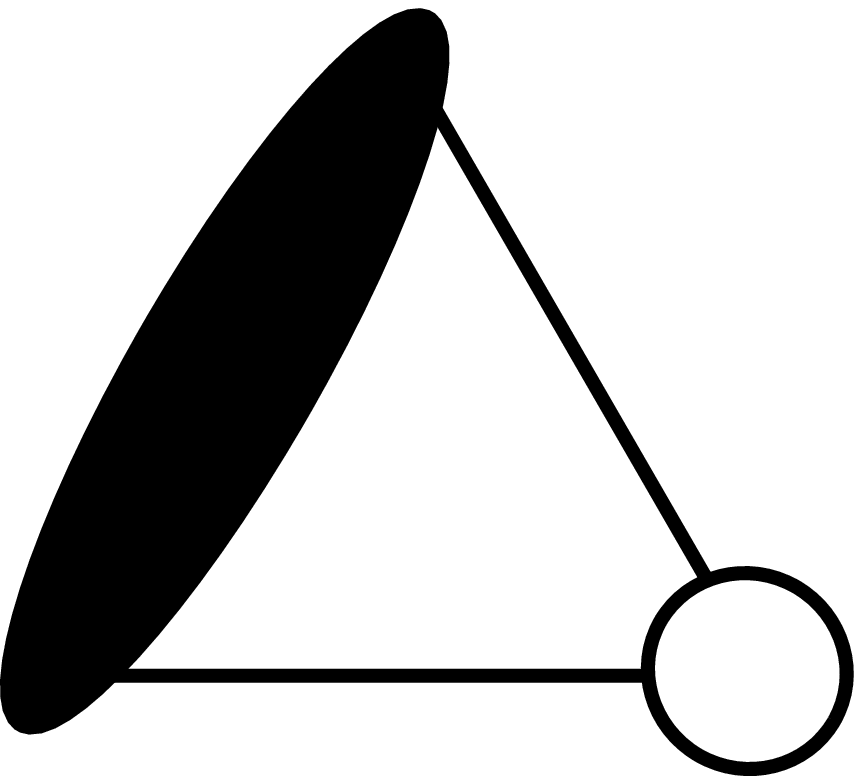}}
\def\tru{\includegraphics[width=2.5mm]{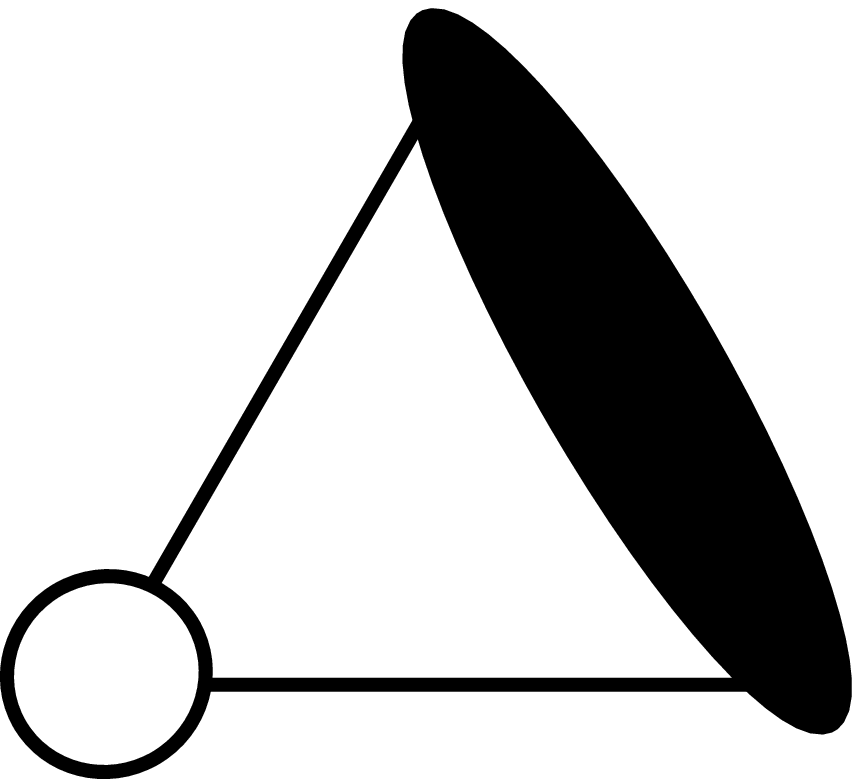}}

\def\itra{\includegraphics[width=2.5mm]{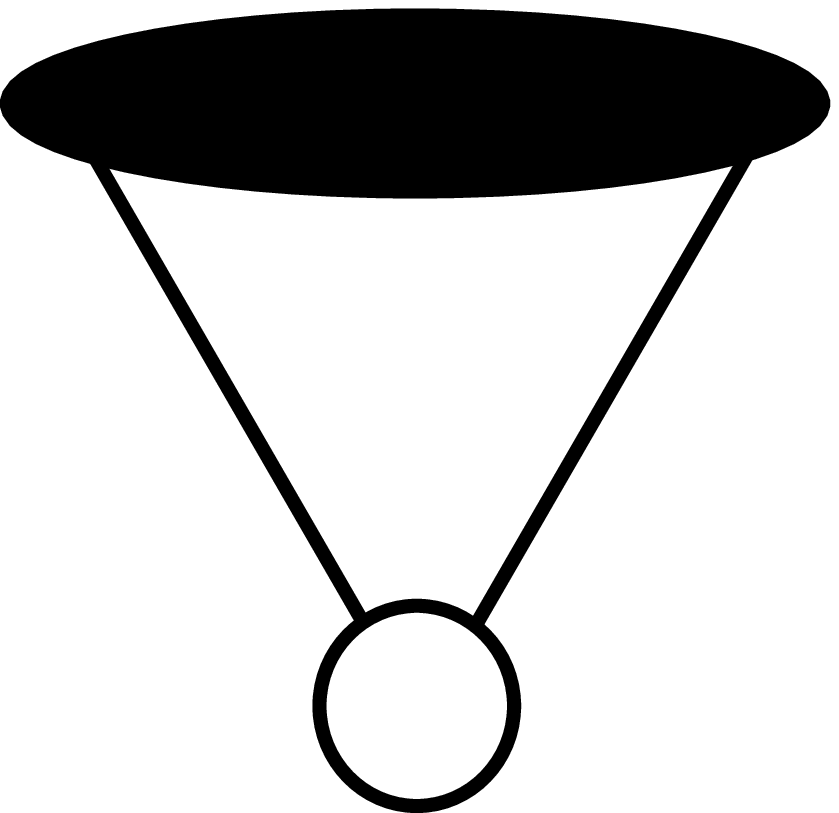}}
\def\itro{\includegraphics[width=2.5mm]{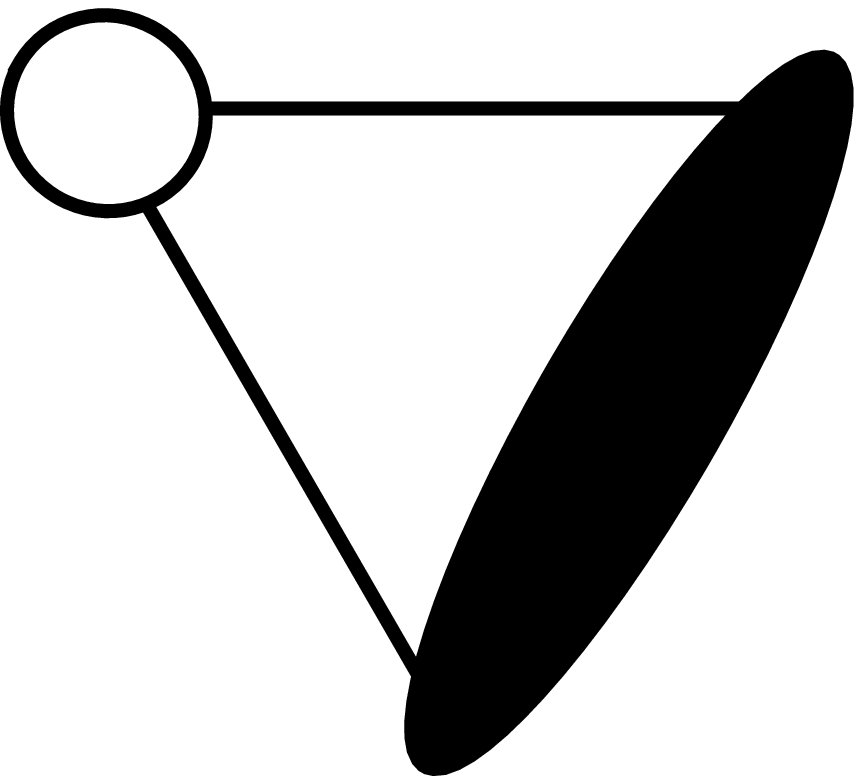}}
\def\itru{\includegraphics[width=2.5mm]{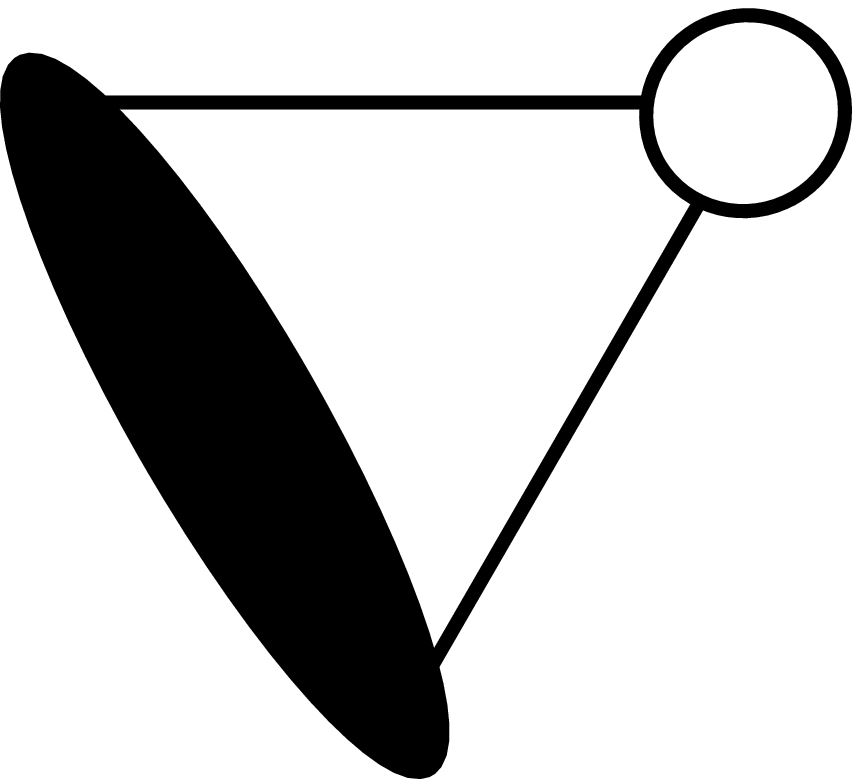}}

\begin{document}

\def\kag{{kagom\'e }}
\def\S{\mbox{\bf S}}
\def\et{{\it et al.}}
\def\tJ{$t{-}J$ }
\bibliographystyle{prsty}

\title{Bond order wave instabilities in doped frustrated antiferromagnets: \\
  ``Valence bond solids'' at fractional filling}
\author{M. Indergand}
\affiliation{
  Theoretische Physik, ETH-H\"onggerberg, CH-8093 Z\"urich
}
\author{A. L\"auchli}
\affiliation{
  Institut Romand de Recherche Num\'erique en Physique des Mat\'eriaux (IRRMA),
  EPFL, CH-1015 Lausanne
}
\author{S. Capponi}
\affiliation{
  Laboratoire de Physique Th\'eorique, CNRS UMR 5152,
  Universit\'e Paul Sabatier, F-31062 Toulouse
}
\author{M. Sigrist}
\affiliation{
  Theoretische Physik, ETH-H\"onggerberg, CH-8093 Z\"urich
}

\date{\today}

\begin{abstract}
We explore both analytically and numerically the properties of doped \tJ models on a class of highly frustrated lattices,
such as the \kag and the pyrochlore lattice.
Focussing on a particular sign of the hopping integral and antiferromagnetic exchange, we find a generic symmetry 
breaking instability towards a twofold degenerate ground state at a fractional filling below half filling. 
These states show modulated bond strengths and only break lattice symmetries. They can be seen as a generalization of the well-known valence bond solid states to fractional filling. 
\end{abstract}
\maketitle
\section{Introduction}

Highly frustrated quantum magnets are fascinating and complex systems where the macroscopic
ground state degeneracy at the classical level leads to many intriguing phenomena at the quantum
level. The ground state properties of spin $S=1/2$ Heisenberg antiferromagnets on the \kag and the
pyrochlore lattice remain still puzzling and controversial in many aspects. 
While the magnetic properties of the Heisenberg and extended models have indeed been studied for quite some time,
the investigation of highly frustrated magnets upon doping with mobile charge 
carriers has started recently~\cite{TsunetsuguKagome,AMLKagome,DidierChecker,BulutKagome,RungeChecker,Fujimoto}. Such interest has been motivated for example by the observation that in some strongly correlated materials, 
such as the spinel compound LiV$_2$O$_4$, itinerant  charge carriers and frustrated magnetic fluctuations interact strongly \cite{KondoLiV2O4}.
Furthermore the possibility of creating optical \kag lattices in the 
context of cold atomic gases has been  pointed out~\cite{SantosKagomeLattice}, making it possible to 
"simulate" interacting fermionic or bosonic models in an artificial setting~\cite{LewensteinKagome}.

At this point we should stress that the behavior in a simple single-band model at weak and at strong 
correlations are not expected to be related in a trivial way. The weak coupling limit allows us to discuss the electronic properties within the picture of itinerant electrons in momentum space based on the notions of a Fermi surface and Fermi surface instabilities
(see e.g.\ Refs.~\onlinecite{TsunetsuguKagome, BulutKagome}). Considering for example the 
Fermi surfaces of a triangular or a \kag lattice at half filling we do not find any obvious
signature of the magnetic frustration present at large $U$. Although at weak coupling 
these systems do not seem to be particularly special, at intermediate to strong coupling
the high density of low-energy fluctuations of the highly frustrated systems display characteristic 
features from which the physics of the frustrated system of localized degrees of freedom will emerge~\cite{AMLKagome,DidierChecker}.

In the following we study a class of highly frustrated lattices, the so called
bisimplex lattices~\cite{BisimplexHenley}, which are composed of corner-sharing 
simplices residing on a bipartite underlying lattice. We restrict ourselves to the
triangle and the tetrahedron as the basic building blocks in the following. This
class hosts lattices such as the \kag or the pyrochlore lattice. 

Our main result is the spontaneous symmetry breaking between the simplices, that are located on different sublattices 
of the underlying bipartite lattice.
This instability takes place at the electron density $n=2/3$ for the triangle based lattices like the \kag lattice and at the density $n=1/2$ for the tetrahedron based lattices like the pyrochlore lattice and leads to a bond order wave (BOW) phase where the kinetic 
energy is staggered for the neighboring simplices. 
This instability is driven by a 
cooperative effect of the kinetic energy {\em and} the exchange interactions.

The outline of the paper is the following:
In Sec.~\ref{sec:model} we introduce the lattices and the model and
provide general arguments for the occurrence of the spontaneous symmetry breaking. These arguments are based on the analysis of the  
spectrum of isolated simplices and on a  
doped quantum dimer model. 
We then report in Sec.~\ref{sec:hf} mean-field calculations for the \kag and the pyrochlore lattices which underline the symmetry breaking tendency. 
In Sec.~\ref{sec:numerics} we present numerical results for the \kag lattice, that were obtained by exact diagonalization (ED) and by the contractor renormalization algorithm (CORE). 
In Sec.~\ref{sec:resultsstrip} the 1D analogue of the \kag lattice, the so-called \kag strip, is treated with two powerful methods that are available for 1D systems: the density matrix renormalization group (DMRG) and a fermionic renormalization group (RG)
and bosonization analysis.
Finally we summarize and conclude in Sec.~\ref{sec:summary}.

\section{General Motivation}
\label{sec:model}
\begin{figure}
  \includegraphics[width=\linewidth]{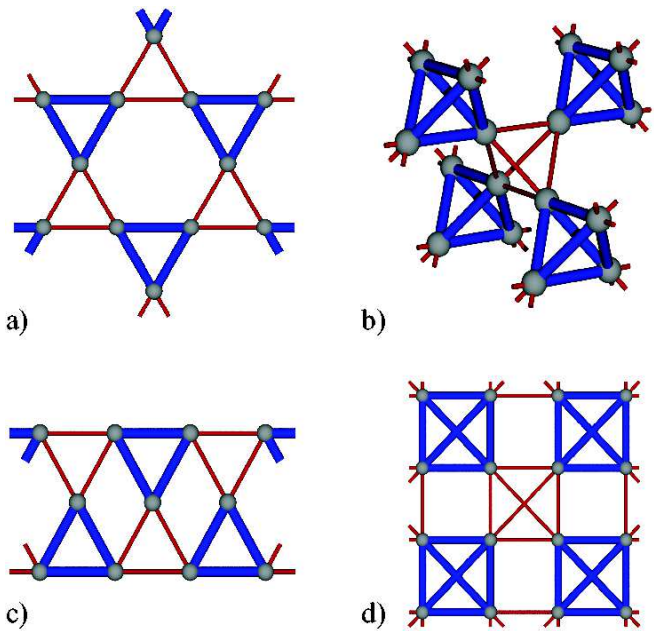}
  \caption{(Color online)
    Four different bisimplex-lattices. 
    The \kag lattice (a) and the pyrochlore lattice (b)
    together with their lower dimensional analogues, the \kag strip (c)
    and the checkerboard lattice (d).
    The two types of corner sharing units (``up'' vs ``down'') 
    are distinguished by the line width. They correspond to the bond
    order wave symmetry breaking pattern occurring at $n=2/3$ on
    the triangle based lattices and at $n=1/2$ on the tetrahedron based
    lattices.
    \label{fig:lattices}
  } 
\end{figure} 
The bisimplex lattices shown in Fig.~\ref{fig:lattices} consist of corner-sharing simplices (triangles or tetrahedra) that are located on an underlying bipartite lattice.
Therefore, we can separate the triangles and the tetrahedra into two different classes, which is visualized in 
Fig.~\ref{fig:lattices} by a different line-style (light and bold bonds).
To refer to the simplices of a given class we call  them  ``up''- and ``down''-simplices, and we use
the same terminology to distinguish the triangles, the tetrahedra or the lattice bonds.

In this section we provide two different arguments why for correlated electrons on such lattices at the fractional filling with two electrons per ``up''-simplex (one electron per simplex)
 a spontaneous symmetry breaking can be expected, resulting in a phase where the ``up''-simplices differ from the ``down''-simplices.
For the \kag and the pyrochlore lattice the inversion symmetry is broken in this phase, whereas for the lower-dimensional analogues, the \kag strip and the checkerboard lattice, translational symmetry is broken.
The following two arguments apply generally to all bisimplex lattices shown in Fig.~\ref{fig:lattices}. They are simple and  illustrate the basic underlying physics. 
In the remaining sections we will provide detailed numerical and analytical evidence for several of these bisimplex lattices, which show that these arguments provide the correct picture.
In the following we use (unless otherwise specified) the \tJ model to describe the correlated electrons on these lattices. The \tJ Hamiltonian is given by
\begin{eqnarray}\label{HtJ}
  \hat{H}_{t{-}J}&=&-t\sum_{\langle ij \rangle}\sum_{\s}\mathcal{P}(c^{\dag}_{i\s}c^{\phantom\dag}_{j\s}+\hc)\mathcal{P}\\
  &&+\ J\sum_{\langle ij \rangle} (\mathbf{S}_i\cdot\mathbf{S}_j-\frac{1}{4}\,n_in_j),\nonumber
\end{eqnarray}
where the restriction to the subspace of singly occupied sites is enforced by the projection operator 
$\mathcal{P}=\prod_{i}(1-n_{i\up}n_{i\down})$. The hopping amplitude $t$ is always
chosen to be positive.
A negative sign of $t$ will most likely induce ferromagnetic tendencies at the fillings we are considering \cite{KoretsuneFerromagnetism}.

\subsection{The limit of decoupled simplices}
\label{sec:aniso}
To get a basic understanding of the effect of doping in highly frustrated lattices we first consider the
limit of decoupled simplices by turning the couplings within the ``down''-simplices off. To connect this limit with the uniform lattice we use the parameter $\alpha\ (0\le \alpha \le 1)$ that tunes the coupling strength of the ``down''-bonds 
as $(\alpha t, \alpha J)$, while the coupling strength on the ``up''-bonds is kept constant at $(t,J)$. 
For $\alpha<1$ the inversion symmetry between the ``up''- and the ``down''-simplices is explicitly broken by the Hamiltonian. 
The eigenvalues of $\hat{H}_{t{-}J}$ and their degeneracies are listed in Table~\ref{tab:TriangleStates} for a single triangle and a 
single tetrahedron. For $t>0$ and $J\ge 0$, there is a non-degenerate 
state with $N_e=2$ that has the lowest energy of all states and is separated from the remaining spectrum by a finite gap. This state has at the same time the lowest kinetic energy ($-2t$ or $-4t$, respectively) of all states and 
gains the maximal exchange energy ($-J$) for two spins. This state is not frustrated anymore because
it minimizes the kinetic and the exchange energy at the same time.
\begin{table}
  \begin{tabular*}{\linewidth}
    {@{\hspace{0.2cm}}l|@{\extracolsep{\fill}}r@{\extracolsep{\fill}}c@{\extracolsep{\fill}}c|@{\extracolsep{\fill}}r@{\extracolsep{\fill}}c@{\hspace{0.1cm}}|}  
    \hline\hline
    &         \multicolumn{3}{l|}{triangle}     &     \multicolumn{2}{l|}{tetrahedron}       \\
    
    $N_e$& Energy  & Degen.      & & Energy     & Degen.  \\  \hline
    0  &     $0$ & $1\times 1$ & & $0$        & $1\times 1$ \\  \hline
    1  &   $-2t$ & $1\times 2$ &*& $-3t$                          & $1\times 2$ \\
       &     $t$ & $2\times 2$ & &   $t$                        & $3\times 2$ \\  \hline
    2  & $-2t-J$ & $1\times 1$ &*& $-4t-J$              & $1\times 1$ \\
       &   $t-J$ & $2\times 1$ & &        $-J$              & $3\times 1$ \\
       &    $-t$ & $2\times 3$ &*&  $2t-J$              & $2\times 1$ \\
       &    $2t$ & $1\times 3$ & & $-2t$              & $3\times 3$ \\  
       &         &             & &  $2t$              & $3\times 3$ \\  \hline
    3  & $-3J/2$ & $2\times 2$ &*&$-2t-3J/2$             & $3\times 2$ \\ 
       &     $0$ & $1\times 4$ & &        $-3J/2$             & $2\times 2$ \\
       &         &             & &  $2t-3J/2$             & $3\times 2$ \\
       &         &             & &  $-t$             & $3\times 4$ \\
       &         &             & &  $3t$             & $1\times 4$ \\  \hline
    4  &         &             & & $-3J$ & $2\times 1$ \\
       &         &             & & $-2J$  & $3\times 3$ \\
       &         &             & & $0$ & $1\times 5$ \\  \hline\hline
  \end{tabular*}
  \caption{
    \label{tab:TriangleStates}
    Classification of the eigenstates of the \tJ model on a  triangle and on a tetrahedron. 
    The degeneracy is given in the form $r\times (2S+1)$, where $r$ is the dimension of 
    the irreducible representation of $\mathcal{S}_3$, respectively $\mathcal{S}_4$, 
    and $S$ is the total spin of the state. The asterix denotes the states retained in the
    CORE calculations for the \kag system, see text.}
\end{table}
After having revealed this particularly stable state with two electrons on a single unit, we know that for $\alpha=0$ the system with two electrons per ``up''-simplex has a non-degenerate ground state and a finite gap to all excitations.
In the tight-binding model the gap decreases with increasing $\alpha$ and vanishes exactly at $\alpha=1$. 
For interacting electrons, however,
the strong correlations that are induced by the constraint in the kinetic term of the \tJ model and the frustration of the exchange term, that arises from the strong frustration of the lattice, are accompanied by a tendency to occupy rather local states (in our case, they still keep a substantial part of their kinetic energy) and to pair up the electrons in nearest neighbor singlets.
Therefore, it is a reasonable possibility that the system even for $\alpha=1$ has a finite gap and breaks the symmetry between the ``up''- and the ``down''-simplices spontaneously in order to profit from the bipartite and non-frustrated structure of the underlying lattice.
Such an instability has the character of a bond order wave -- i.e., modulated expectation
values of the bond energies --
and yields an insulating state which breaks inversion or translational symmetry, respectively. 

\subsection{Doped quantum dimer model}
\label{sec:qdm}
In the previous sections we considered the case $\alpha=0$, where the Hamiltonian itself is not invariant under inversion symmetry. In order to get some insight into the mechanism of spontaneous symmetry breaking, it is desirable to treat ``up''- and ``down''-simplices on equal footing. In the following, we present a 
simple but illustrative model with a symmetric Hamiltonian that breaks spontaneously the symmetry between the ``up''- and ``down''-simplices in the ground state. 

A close inspection of the wavefunction of the lowest energy eigenstate of two electrons on either a triangle or a tetrahedron
reveals that it consists of the equal amplitude superposition of all possible positions of the singlet formed by the two electrons:
\begin{equation}
|\psi_{GS}\rangle=\frac{1}{\mathcal{N}}\sum_{i < j} (c^\dagger_{i,\uparrow}c^\dagger_{j,\downarrow}-c^\dagger_{i,\downarrow}c^\dagger_{j,\uparrow})|0\rangle,
\end{equation}
where the normalization $\mathcal{N}=\sqrt{3}$ for the triangle and  $\mathcal{N}=\sqrt{6}$ for the tetrahedron. This
wavefunction motivates us to design a simple quantum dimer model which on each triangle prefers the exact wavefunction
described above. Such a Hamiltonian reads for example for the \kag lattice:
 \begin{eqnarray}
 \label{eqn:qdm}
 H_{\mathrm{QDM}}&=&-t \sum_{\triangle}\big[|\tro\rangle\langle\tru|+|\tru\rangle\langle\tra|+|\tra\rangle\langle\tro|+\hc \big] \nonumber\\
 	&&-t\sum_{\bigtriangledown}\big[|\itro\rangle\langle\itru|+|\itru\rangle\langle\itra|+|\itra\rangle\langle\itro|+\hc \big]
 \end{eqnarray}
where the Hilbert space consists of all coverings of the \kag lattice with $N_c$ nearest-neighbor dimers and $N_c$ monomers, 
$N_c$ counting the number of unit cells. This corresponds to the situation at $n=2/3$ in the \tJ model. The interpretation is simple:
the antiferromagnetic exchange term favors all the electrons to pair up into singlets, while the kinetic energy term delocalizes the singlets on a triangle.
The quantum dimer model for the tetrahedron based lattices are defined by letting a single singlet resonate on a tetrahedron.
This simple model allows us to find the exact ground state on these lattices. The ground state is twofold degenerate and each state 
is the direct product of equal amplitude resonances on the same type of triangles/tetrahedra, either all
``up'' or all ``down''. In such a situation each resonating dimer can independently fully optimize its kinetic energy.
The argument has much in common with the reasoning for the close packed dimer model on the pyrochlore lattice discussed in Ref.~\onlinecite{MoessnerSpNPyro}.

Although this model is only a cartoon version of the real electronic system, it illustrates how the tendency of the electrons to form nearest-neighbor singlets 
obstructs the motion of the singlets between corner-sharing simplices, but within a given simplex an individual singlet can hop without obstacles and optimize its kinetic energy.
The bipartite nature of the underlying lattice allows for the localization of the singlets on simplices  without interference and triggers in this way the spontaneous symmetry breaking. 

\section{Mean-field Discussion}
\label{sec:hf}
In this section we present a mean-field calculation for the \kag lattice and the pyrochlore lattice.
 The mean-field discussion is particularly valuable for the pyrochlore lattice, as due to its higher dimension it is less affected by fluctuations and is presently not treatable with numerical methods.
We can show that in the mean-field analysis the spontaneous inversion symmetry breaking, discussed in the previous sections, is also the natural and leading instability. 

We start with discussing the properties of the nearest-neighbor tight-binding model on the \kag and the pyrochlore lattice, given by 
\begin{equation}\label{H0}
\hat{H}_{0}=-\mu\hat{N}-t\sum_{\br, \s}\sum_{m\neq n}\sum_{\nu=\pm1}c^{\dag}_{\br+\nu\ba_m,\s}c^{\phantom\dag}_{\br+\nu\ba_n,\s},
\end{equation}
where $\s\in\{\up,\down\}$ is the spin index and the indices $m,n$ run from zero to the dimension of the lattice, $d$. Furthermore, $\br$ is  an elementary lattice vector connecting unit cells  and the vectors $\ba_0,\dots,\ba_d$ point to the vertices of an elementary triangle (tetrahedron) in the \kag (pyrochlore) lattice, $\ba_0=0$.     
$\hat{N}$ is the total electron number operator and $\mu$ is the chemical potential.
In the following we will use units where $t=1$ and we will always choose $\mu=-1$ for the \kag and $\mu=-2$ for the pyrochlore lattice, which corresponds to two electrons per unit cell.
$\hat{H}_0$ can be diagonalized in reciprocal space and can be written as  
\begin{equation}\label{diagH0}
\hat{H}_{0}=\sum_{\bk, m, \s}\xi^{\phantom\dag}_{\bk m}\g^{\dag}_{\bk m\s}\g^{\phantom\dag}_{\bk m\s},
\end{equation}
with
\begin{equation}
-\xi_{\bk 0}=\xi_{\bk 1}\equiv \xi_{\bk}, \quad \xi_{\bk m}=d+1 \ \textrm{for}\ m>1.
\end{equation}
For the \kag lattice we have 
\begin{equation}\label{xiKagome}
\xi_{\bk}=\sqrt{1+8\cos(k_1/2)\cos(k_2/2)\cos([k_1-k_2]/2)}
\end{equation}
with $k_m=\bk\cdot\ba_m$ and  we 
refer to Ref.~\onlinecite{isoda} for an explicit formula for $\xi_{\bk}$ for the pyrochlore lattice.

\begin{figure}[htbp]
      \includegraphics[width=\linewidth]{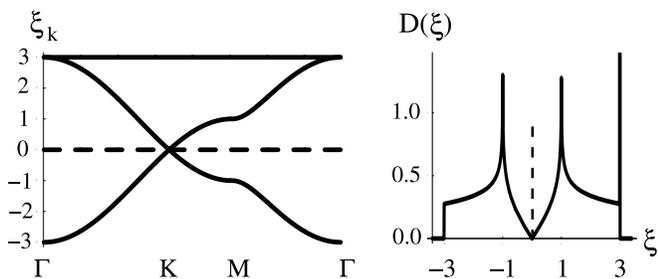}
        \caption{The \kag bands and the density of states per unit cell and spin. The energy is measured in units of $t$. \label{kagomebands}} 
\end{figure}

 The three bands of the \kag lattice consist of one flat band and two dispersing bands. The dispersing bands are identical to the bands of a honeycomb lattice. They are shown together with the density of states per unit cell and spin in Fig.~\ref{kagomebands}.  
Note, that around the points K and $-\textrm{K}$  the dispersion shows a Dirac spectrum, i.e., the bands $\xi_{\bk 0}$ and $\xi_{\bk 1}$
 touch at these points with linear dispersion. For the given chemical potential the Fermi surface reduces to points at K and $-\textrm{K}$ and the density of states vanishes linearly with $\xi$, i.e., we have $D(\xi)\propto |\xi|$ for small $\xi$.

\begin{figure}[htbp]
      \includegraphics[width=\linewidth]{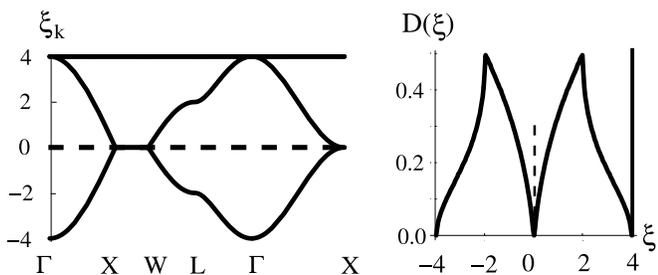}
        \caption{The pyrochlore bands and the density of states per unit cell and spin. The energy is measured in units of $t$. \label{pyrochlorebands}} 
\end{figure}

 The four bands of the pyrochlore lattice consist of two flat bands and two dispersing bands. The dispersing bands are identical to the bands of a diamond lattice. They are shown together with the density of states per unit cell and spin in Fig.~\ref{pyrochlorebands}.
Note, that $\xi_{\bk}$ vanishes along the lines connecting X and W. The density of states also vanishes  linearly at zero up to logarithmic corrections, i.e., we have $D(\xi)\propto |\xi|\log(|\xi|)$ for small $\xi$.

Systems  with this form of the  density of states at the Fermi level are neither band-insulators nor normal metals, therefore, they are sometimes called semi-metals or zero-gap semiconductors.
Although they have an even number of electrons per unit cell and no fractionally filled bands, they have no energy gap at the Fermi surface. Fermi surface instabilities are suppressed in this situation. 
There is no Cooper-instability that leads to an obvious breakdown of perturbation theory for arbitrarily small attractive interactions, as the particle-particle polarization function involves the convergent integral $\int d\xi D(\xi)/2|\xi|$ at zero temperature.    
For the half filled honeycomb lattice it has been shown that the Coulomb  interactions lead to non-Fermi liquid behavior and that strong enough Coulomb interactions lead to antiferromagnetic order and to the opening of a charge gap\cite{SorellaHoneycomb,martelo,gonzalez,gonzalez2}.
  The situation in the \kag and the pyrochlore lattice at the filling considered here is different.
Because the lattices are not at half filling, it is not obvious that even arbitrarily large $U$ would enforce a charge gap (Mott insulator) and an antiferromagnetic order would be hampered by the frustrated topology of the lattice.
However, if we consider the triangles (tetrahedra) as the fundamental units of our lattice we obtain the honeycomb (diamond) lattice and the properties of this underlying bipartite lattice will be reflected in the ground state and provide a way to circumvent the frustration effects.

Our goal is  to study the electron-electron interactions of the $\hat{H}_{t{-}J{-}V}$ Hamiltonian given by
\begin{equation}\label{HtJV}
\hat{H}_{t{-}J{-}V}=\mathcal{P}\hat{H}_0\mathcal{P}+\sum_{\langle ij\rangle}J\,\mathbf{S}_i\cdot\mathbf{S}_j+Vn_in_j,
\end{equation}
We will show that both the exchange and the repulsion term favor the bond order wave instability.
As the projection operator $\mathcal{P}$ is difficult to handle in analytic calculations, the projection is often approximated by a purely statistical renormalization of the Hamiltonian with Gutzwiller factors \cite{Gutzwiller}. We obtain a renormalized Hamiltonian without constraints given by
 \begin{equation}\label{Hr}
\hat{H}_{\mathrm{r}}=g_t\,\hat{H}_0+\sum_{\langle ij\rangle} Jg_J\,\mathbf{S}_i\cdot\mathbf{S}_j+Vn_in_j.
\end{equation}
The renormalization is given by the Gutzwiller factors $g_t=2\d/(1+\d)$ and $g_J=4/(1+\d)^2$ and $\d$ is the hole doping measured from half filling. Note, that the nearest-neighbor repulsion is not renormalized by a statistical factor.

In the following we will determine the critical $J$ and $V$ for spontaneous symmetry-breaking in this model within mean field theory.
Superconductivity is a possible way of spontaneous symmetry breaking. As it is an instability in the particle-particle channel, the relation $\xi_{\bk}=\xi_{-\bk}$, which is ensured by inversion and time reversal symmetry, plays an essential role.
Concerning the symmetry of the order parameter, we can restrict ourselves to singlet pairing in the spin sector, because the nearest neighbor interaction is antiferromagnetic, and to $s$-wave pairing in the orbital sector, because in this way we obtain a nodeless, even gap-function. 

Another possibility of spontaneous symmetry breaking is an instability in the particle-hole channel.
Such instabilities tend to occur if a nesting condition of the form $\x_{\bk}=-\xi_{\bk+\mathbf{q}}$ is fulfilled. In general, this condition is not ensured by basic symmetries and therefore instabilities in the particle-hole channel are much more special than superconducting instabilities.
In our case, the relation $\xi_{\bk 0}=-\xi_{\bk 1}$ can be considered as perfect nesting with $\mathbf{q}=0$.
Therefore, the relevant question is which one of the two considered instabilities is dominant in our system.
In order to answer this question we consider the following single-particle Hamiltonian, 
\begin{equation}\label{Htrial}
\hat{H}_{\mathrm{trial}}=\hat{H}_{0}+\D_{\mph}\hat{H}_{\mph}+\D_{\mpp}\hat{H}_{\mpp}
\end{equation}
where we have introduced the two quadratic Hamiltonians 
\begin{eqnarray}\label{HppHph}
\hat{H}_{\mph}&=&\sum_{\br,\s}\sum_{m\neq n}\sum_{\nu=\pm1}\nu \,c^{\dag}_{\br+\nu\ba_ m,\s}c^{\phantom\dag}_{\br+\nu\ba_n, \s},\\
\hat{H}_{\mpp}&=&\sum_{\br}\sum_{ m\neq n}\sum_{\nu=\pm1}(c_{\br+\nu\ba_m,\down}c_{\br+\nu\ba_n,\up}+\hc).\nonumber
\end{eqnarray}
The idea is to calculate the expectation value of $\hat{H}_{\mathrm{r}}$ for the ground-state of $H_{\mathrm{trial}}$ and to choose the variational parameters $\D_{\mpp}$ and $\D_{\mph}$ such that this expectation value is minimized.
In terms of the operators that diagonalize $\hat{H}_0$ we can express the pairing operators
as
\begin{eqnarray}\label{HppHphpairing}
\hat{H}_{\mph}&=&\sum_{\bk,\s}ix_{\bk}\g^{\dag}_{\bk 0\s}\g^{\phantom\dag}_{\bk 1 \s}+\hc,\\
\hat{H}_{\mpp}&=&\sum_{\bk, m}\epsilon_{\bk m}\g_{-\bk m\up}\g_{\bk m\down}+\hc.\nonumber
\end{eqnarray}
with the relations
\begin{equation}\label{epsilonx}
\xi_{\bk m}=\e_{\bk m}-\mu, \qquad x_{\bk}^2=\xi_0^2-\xi_{\bk}^2.
\end{equation}
For small values of $\D_{\mph}$ and $\D_{\mpp}$ we can expand the ground-state expectation value of $H_{\mathrm{r}}$ in terms of $\D_{\mpp}^2$ and $\D_{\mph}^2$.
Using the Wick theorem, we obtain up to higher order terms
\begin{eqnarray}\label{Hrexp}
\frac{\D E}{bN}&=&\D_{\mph}^2I_{\mph}\left[\tilde{t}-\frac{3\tilde{J}}{8}(I_{\mph}-\chi)-\frac{V}{2}(I_{\mph}-\chi)\right]\\
&+&\D_{\mpp}^2I_{\mpp}\left[\tilde{t}-\frac{3\tilde{J}}{8}(I_{\mpp}-\chi)+\frac{V}{2}(I_{\mpp}+\chi)\right]\nonumber
\end{eqnarray}
where $\D E$ is the deviation from the ground-state expectation value with $\D_{\mph}=\D_{\mpp}=0$.
$N$ is the number of unit cells, $\tilde{t}=tg_t$, $\tilde{J}=Jg_J$, $b$ is the number of bonds in the unit cell  and
\begin{eqnarray}\label{chiIphIpp}
\chi&=&\frac{1}{b}\int_{\xi<0}(-\xi-\mu)D(\xi)d\xi,\\
I_{\mph}&=&\frac{1}{b}\int_{\xi<0}\frac{\xi_0^2-\xi^2}{|\xi|}D(\xi)d\xi,\nn
I_{\mpp}&=&\frac{1}{b}\int\frac{(\xi+\mu)^2}{2|\xi|}D(\xi)d\xi.\nonumber
\end{eqnarray}
Note that only the density of states enters these formulas because we are restricting ourselves to $\mathbf{q}=0$ instabilities.
The system spontaneously breaks inversion ($U(1)$) symmetry, if the coefficient of $\D_{\mph}^2$ ($\D_{\mpp}^2$) in Eq.~(\ref{Hrexp}) changes sign.  If we assume that only one of the parameters $V$ and $J$ is nonzero, we obtain the following expressions for the critical values:
\begin{eqnarray}
J_c^{\mph}=\frac{8g_tt}{3g_J(I_{\mph}-\chi)}&\qquad& V_c^{\mph}=\frac{2g_tt}{(I_{\mph}-\chi)}\\
J_c^{\mpp}=\frac{8g_tt}{3g_J(I_{\mpp}-\chi)}&\qquad& V_c^{\mpp}=\frac{-2g_tt}{(I_{\mpp}+\chi)}
\end{eqnarray}
The numerical values for $J_c$ and $V_c$ are given in TABLE~\ref{parameters}.
\begin{table}[htbp]
  \centering
\begin{tabular*}{\linewidth}{@{\hspace{0.2cm}}c@{\extracolsep{\fill}}c@{\extracolsep{\fill}}c@{\extracolsep{\fill}}c@{\extracolsep{\fill}}c@{\extracolsep{\fill}}c@{\extracolsep{\fill}}c@{\extracolsep{\fill}}c@{\hspace{0.2cm}}}
\hline
\hline
 &$d$&$\mu$&$ b$&$\xi_0$&$\d $&$g_t$&$g_J$  \\
\hline
K&2&-1 & 6& 3   &1/3&1/2&9/4  \\
P&3&-2 &12& 4   &1/2&2/3&16/9 \\
\hline
\end{tabular*}
\begin{tabular*}{\linewidth}{@{\hspace{0.2cm}}c@{\extracolsep{\fill}}c@{\extracolsep{\fill}}c@{\extracolsep{\fill}}c@{\extracolsep{\fill}}c@{\extracolsep{\fill}}c@{\extracolsep{\fill}}c@{\extracolsep{\fill}}c@{\hspace{0.2cm}}}
\hline
 &$\chi$&$I_{\mph}$&$I_{\mpp}$&$J_c^{\mph}$&$J_c^{\mpp}$&$V_c^{\mph}$&$V_c^{\mpp}$  \\
\hline
K&0.43  & 1.08     & 0.59     & 0.91       & 3.58       & 1.53       & -0.98  \\
P&0.32  & 1.05     & 0.62     & 1.36       & 3.33       & 1.81       & -1.43  \\
\hline
\hline
\end{tabular*}
 \caption{\label{parameters} The parameters for the \kag (K) and the pyrochlore (P) lattice. The critical values are given in units of~$t$.
The coefficient of $\D_{\mph}$ (bond order wave) in Eq.~\eqref{Hrexp} is negative for $J>J_c^{\mph}$ ($V=0$) or for {\it repulsive} $V>V_c^{\mph}$ ($J=0$).
The coefficient of $\D_{\mpp}$ (superconductivity) is negative for $J>J_c^{\mpp}$ ($V=0$) or for {\it attractive} $V<V_c^{\mpp}$ ($J=0$).
} 
\end{table}
One can see, that the tendency for inversion symmetry breaking is much stronger than the tendency for superconductivity in both lattices and that both the antiferromagnetic $J$ and the repulsive $V$ support the inversion symmetry breaking.
The integral  $I_{\mph}$ is large  because  the factor $\xi_0^2-\xi^2$ takes its maximum at $\xi=0$ whereas the factor $(\xi+\mu)^2$ in the integral $I_{\mpp}$ is much smaller for small values of $\xi$.    
In other words, superconductivity has the handicap that the potential is proportional to the dispersion $\e_{\bk}$, therefore it is small at the Fermi surface and is only finite due to the finite value of $\mu$. 
The nearest-neighbor repulsion is harmful for Cooper (particle-particle) pairing, as can be seen from Table~\ref{parameters}.
In the particle-hole instability, however, two particles tend to form a singlet on every second triangle (tetrahedron) on the \kag (pyrochlore) lattice. In this way the singlet is still mobile and keeps $-dt$  of its kinetic energy and at the same time reduces the nearest-neighbor repulsion energy from $4V/3$ ($3V/2$) to $V$ on every second triangle (tetrahedron). On the triangles (tetrahedra) without a singlet, the expectation value of the nearest-neighbor repulsion is however still $4V/3$ ($3V/2$).   
In the limit where the kinetic energy is negligible ($t\ll V,J$) also other phases may appear. It is therefore important to emphasize that
a finite kinetic energy is necessary to stabilize the bond order wave, because this phase arises due to the interplay between the kinetic 
and interaction energy. 
 
The  limit of large $V$ was recently discussed in the context of LiV$_2$O$_4$ by Yushankhai \et\ in Ref.~\onlinecite{yushankhai} for the pyrochlore lattice with $n=1/2$. The possibility of inversion symmetry breaking was not considered in that study. But if $V$ is of the order of $t$, the optimization of the kinetic and the repulsion energy can
lead to a compromise which breaks the inversion symmetry.

In the bond order wave phase that we found in this section for the \kag and the pyrochlore lattice, the ``up''-triangles (tetrahedra) have a higher expectation value of the 
kinetic energy than the ``down''-triangles (tetrahedra). Furthermore a gap proportional to $\D_{\mph}$ opens at the Fermi surface. Therefore, the system made a transition
from a semi-metal to an insulator. This transition is similar to the Peierls metal-insulator transition, where a half filled system lowers spontaneously its crystal symmetry 
in order to open a gap at the Fermi surface. Phonons or the elasticity of the crystal play a crucial role in the Peierls transition. In our case, as we showed, the 
transition can be driven by a purely electronic mechanism in an infinitely rigid lattice. In reality, the crystal structure will always relax and in this way additionally enhance 
the transition.                                                                                        

\section{Numerical Results for the \kag lattice}
\label{sec:numerics}

In this section we compare the predictions obtained
in the preceding sections to numerical results for the \kag lattice.
We will first discuss some exact diagonalization (ED) results for
the \kag lattice at $n=2/3$, which confirm the BOW clearly and then we will study the dependence of the excitation gaps on the parameter $\alpha$ using the contractor renormalization (CORE) method. 
In essence the numerical results corroborate the analytical predictions on the
presence of a bond order wave instability for the \kag lattice.

\subsection{Exact Diagonalization results}
\label{sec:ED}
\begin{figure}
  \includegraphics[width=0.95\linewidth]{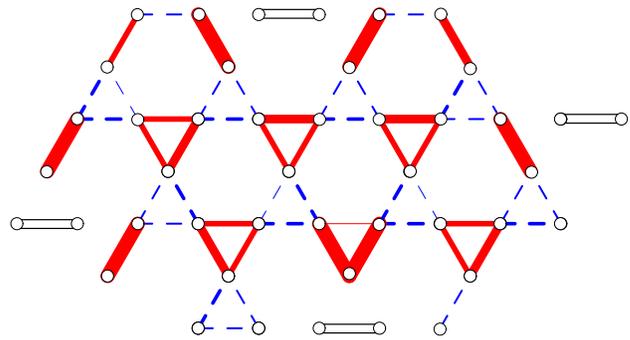}
  \caption{
    (Color online)
    Correlation function of the kinetic energy (Eq.~\protect{\ref{eqn:kinetic_correlation}})
    of a 21 sites \kag sample at $n=2/3$ and $J/t=0.4$. The black, empty bonds denote 
    the same
    reference bond, the red, full bonds negative and the blue, dashed bonds positive
    correlations. The line strength is proportional to the magnitude of the correlations.
    \label{fig:KagomeKineticCorrsED}} 
\end{figure}
The analytical arguments presented in Sec.~\ref{sec:model}
and Sec.~\ref{sec:hf} predict a bond order wave instability at
filling $n=2/3$. In finite, periodic systems this instability
can be detected with a correlation function of the bond
strength, either of the kinetic term or the exchange term.
Here we chose to work with the kinetic term, but the exchange term
gives similar results.
The correlation function is defined as:
\begin{eqnarray}
  C_\mathrm{Kin}[(i,j),(k,l)]&=&\langle\mbox{Kin}(i,j)\ \mbox{Kin}(k,l) \rangle\nonumber\\
  && -\langle\mbox{Kin}(i,j)\rangle\ \langle\mbox{Kin}(k,l)\rangle,
  \label{eqn:kinetic_correlation}
\end{eqnarray}
where
\begin{equation}
\mbox{Kin}(i,j)=-\ \sum_{\sigma}c^\dagger_{i,\sigma}c_{j,\sigma}+\hc,
\end{equation}
and $(i,j)$ and $(k,l)$ denote two different nearest-neighbor bonds of the
\kag lattice, that have no common site. This correlation function has been calculated for all distances
in the ground state of a finite \kag sample with 21 sites, containing 7 holes,
at $J/t=0.4$. The result is plotted in Fig.~\ref{fig:KagomeKineticCorrsED}.
The reference bond uniquely belongs to a certain class of triangles (``up''-triangle in
our case). Based on the theoretical picture one expects the 
correlation function to be positive for all bonds on the same type of triangles
and negative on the others. This is indeed what is seen in 
Fig.~\ref{fig:KagomeKineticCorrsED}. We have also calculated the same quantity for 
$J/t=1$ and $J/t=2$ and the bond order wave correlations (not shown) were becoming
even stronger for larger $J/t$. In this respect the ED calculations confirm
the qualitative picture developed above, that the homogeneous $t{-}J$ model
on the \kag lattice at $n=2/3$ has an intrinsic instability towards a spontaneous 
breaking of the inversion symmetry.

\subsection{CORE results}
We know from Sec.~\ref{sec:aniso}, where we studied the limit of decoupled units, 
that the system has a finite gap in all sectors for $\alpha=0$.
In this section we show  how the gaps of the different sectors of excitations depend on the parameter $\alpha$.

The CORE algorithm~\cite{Morningstar1996,Capponi2004,auerbach}, which is based on the existence of strong subunits (triangles) that are only weakly coupled, provides a suitable method to perform such a study.
 This is certainly true for small values of $\alpha$ where the system is naturally divided into weakly coupled subunits. For $\alpha=1$ the CORE algorithm imposes a certain bias, as ``up''- and ``down''-triangles are not treated on equal footing, however, we have good evidence from ED and from the mean field analysis, that the system itself has a strong tendency to break inversion symmetry spontaneously and therefore we can expect the CORE results to be reliable even for $\alpha=1$.
In fact, we compare the CORE results to the ED results for the smaller clusters, and the good agreement shows that the finite excitation gaps at $\alpha=1$ are not an artefact of the method. 

The CORE method extends the range of tractable sizes of finite clusters,
 based on a careful selection of relevant 
low-energy degrees of freedom. In order to apply this algorithm, the lattice has to be divided into
blocks; here, we naturally choose the ``up''-triangles. A reduced Hilbert space is defined by
retaining a certain number of low-lying states on each block.  The choice of the states to keep
depends also on the quantities to be obtained. While for a ground state calculation fewer 
states already provide good results, one has to retain usually more states to calculate the excited states.
Here we choose to keep the 4 lowest states in the 3-electrons 
sector, the 7 lowest states  with 2 electrons and the 2 lowest states in the 1-electron sector. These states are denoted with an asterix  in 
Table~\ref{tab:TriangleStates}. This choice leads to a reduction of the local triangle basis from 
27 down to 13 states, thus allowing indeed to perform simulations on larger lattices
than would be possible by conventional ED.

Then, by computing the exact low-lying eigenstates of two coupled triangles,
we calculate the effective interactions at interaction range two for each value
of $\alpha$ and we neglect longer range terms. Comparison to ED data on the smaller clusters
shows that this approach gives very good results.

The basic excitation gaps of interest in the present problem are the spin gap, the
single particle gap and the two particle gap. These are defined as follows:
\begin{equation}
  \Delta_{S=1}=E(N_e,1)-E(N_e,0),
  \label{eqn:SpinGap}
\end{equation}
\begin{equation}
  \Delta_{1p}=\frac{1}{2}[E(N_e+1,1/2)+E(N_e-1,1/2)]-E(N_e,0),
  \label{eqn:1Pgap}
\end{equation}
\begin{equation}
  \Delta_{2p}=\frac{1}{2}[E(N_e+2,0)+E(N_e-2,0)]-E(N_e,0),
  \label{eqn:2Pgap}
\end{equation}
where $E(N_e,S^z)$ denotes the ground state energy in the sector with $N_e$ electrons
and spin polarization $S^z$.

\begin{figure}
  \includegraphics[width=\linewidth]{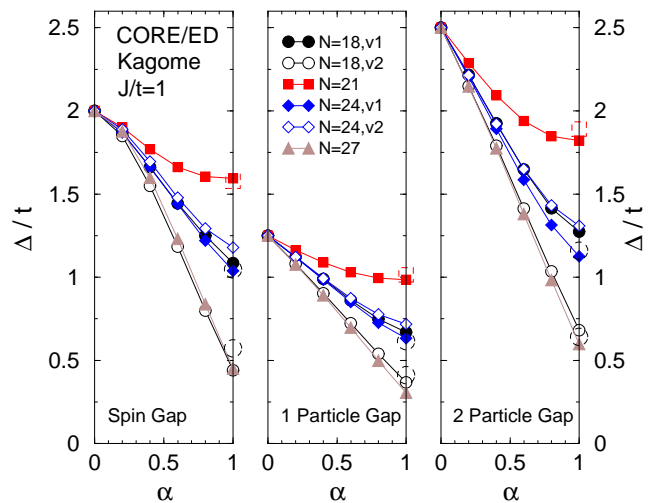}
  \caption{
  (Color online)
  Excitation gaps of the \tJ model on the \kag lattice at $J/t=1$ as a
  function of the parameter~$\alpha$, which denotes the ratio of the intertriangle
  to the intratriangle couplings. The gaps are obtained by the CORE method for
  different sample sizes (and geometries). For the smaller samples (18,v1;\,18,v2;\,21) ED data is shown
  for comparison at $\alpha=1$.
  \label{fig:CoreKagomeGaps}} 
\end{figure}

We have determined these gaps on \kag finite size samples at $n=2/3$ and $J/t=1$ containing 18 to 27 sites.
Two different versions of samples with 18 and 24 sites have been treated (v1 and v2).
The results are displayed in Fig.~\ref{fig:CoreKagomeGaps}. There are two main observations: (1) the gaps
do not close for any $\alpha \in [0,1]$, giving additional evidence for the proposed symmetry breaking; (2)
there is a strong dependence of the gap curves on the specific sample. Note that there is no
discrepancy between ED and CORE results. The second phenomenon can be understood from the discretization of the
finite size Brillouin zones~: indeed, the measured gaps directly depend on the distance between  the closest point in the Brillouin
zone to the corner of the zone, the $K$-point, which is the point where the gap opens in the mean-field picture (Fig.~\ref{kagomebands}). 
 The $18,$v2 and the 27 sites samples both contain this specific point
and differ only slightly in the values of the gap. Thus, our numerical data
support the claim of a finite gap for all $\alpha \in [0,1]$.
The strong dependence is at the same time also a hint towards a sizable dispersion of the excitations  in this system.

\begin{figure*}[t]
  {\large (a) $-\sum_{\sigma}\langle c^\dagger_{i,\sigma}c^{\phantom\dagger}_{j,\sigma}\rangle$}
  \includegraphics*[width=0.95\linewidth]{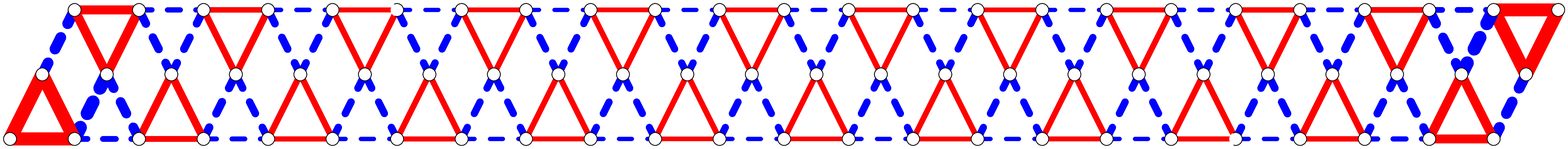}
  $\mbox{}$\\
  \vspace{5mm}
  {\large (b) $\langle\mathbf{S}_i\cdot\mathbf{S}_j\rangle$}
  \includegraphics*[width=0.95\linewidth]{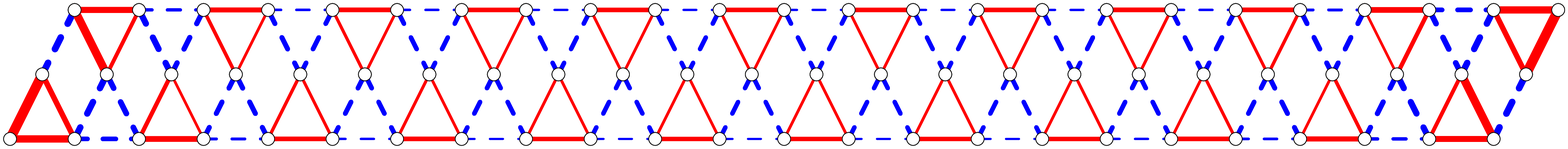}
  \caption{
    (Color online)
    DMRG results for a $L=24$ \kag ladder at $J/t=0.4$ and $n=2/3$.
    (a) Local bond strength deviation of the kinetic term.
    Red, full bonds are stronger (lower in energy) than the average kinetic energy per bond.
    Blue, dashed bonds are weaker than average bonds.
    (b) Local bond strength deviation of the exchange term.
    The color pattern are the same as in the upper panel.
    The thickness of the bond denotes the deviation from the average value per bond.
    Note that the pattern of the kinetic and the exchange term are {\em in phase}.
    \label{fig:BondStrength}
  }
\end{figure*}

\section{DMRG and RG Results for the \kag strip}
\label{sec:resultsstrip}

In this section we study with two different methods the \kag strip, shown in Fig.~\ref{fig:lattices} c),
 as the 1D analogue of the \kag lattice.
This lattice has been introduced in Ref.~\onlinecite{OneDKagomeMagnetic}, where it was shown to share
some of the peculiar magnetic properties of the 2D \kag lattice.
We report extensive
density matrix renormalization group (DMRG) calculations
\cite{dmrg_white}, both for the $t{-}J$ and the Hubbard model.
Furthermore, we apply the renormalization group (RG) and the bosonization techniques to the weak coupling Hubbard model.
Our results show, that for both, the Hubbard and the \tJ model, and irrespective of the coupling strength the BOW phase is realized in the \kag strip at $n=2/3$.  

\subsection{DMRG results}

The \kag strip - being 1D - offers the opportunity to perform DMRG simulations,
thus allowing a rather detailed numerical study of large systems.
We first discuss the properties of the $t{-}J$ model at $n=2/3$ on this lattice,
and then make a connection by investigating the Hubbard model at
different values of $U$ to the analytical weak-coupling results of the following section. For both models we report numerical evidence for the
presence of the bond order wave instability for a large range of interaction strengths.

\begin{figure}
  \includegraphics[width=0.9\linewidth]{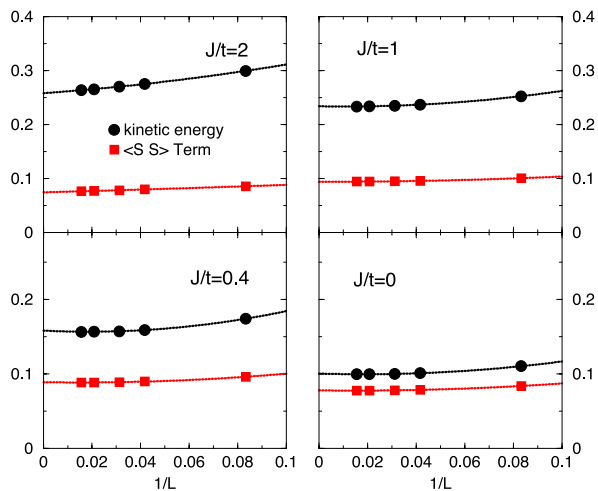}
  \caption{(Color online) DMRG results for the alternation of the bond strength 
   of the kinetic term and the spin exchange term as a function of inverse system size $1/L$,
   for different values of $J/t$. \label{fig:dmrgorderparameters}} 
\end{figure}
In contrast to the periodic systems considered above within ED, the DMRG 
works most efficiently for open boundary conditions (OBC). In the present
context this has the additional advantage that for even length $L$ of the strip
only one of the two degenerate ground states is favored, and we can directly measure
the local bond strength.
For the purpose of illustration we show the local bond strengths for a system of
$L=24$ in Fig.~\ref{fig:BondStrength}. The upper panel shows the difference of
the local kinetic energy with respect to the average, while the lower panel shows
the local expectation value of the spin exchange term, using the same convention.

The calculated pattern resembles the schematic picture drawn in Fig~\ref{fig:lattices}~c).
In order to address the behavior in the thermodynamic limit we measure the bond strength alternation,
i.e., the difference between the expectation values of the operators $\mathrm{Kin}(i,j)$ and $\mathbf{S}_i\cdot\mathbf{S}_j$ in the middle of the system for different 
lengths $L$ and values of $J/t$. The scaling of these quantities is shown in 
Fig.~\ref{fig:dmrgorderparameters}. The finite size corrections are rather small and all the
order parameters extrapolate to finite values, irrespective of the value of $J/t$. 
Note that even for the case $J/t=0$ there is both a finite alternation of the kinetic energy
{\em and} the magnetic exchange term. The alternation of the magnetic exchange energy
is roughly the same for all values of $J/t$. The alternation of the kinetic energy however is
increased with growing $J/t$ ratio.

\begin{figure}
  \includegraphics[width=\linewidth]{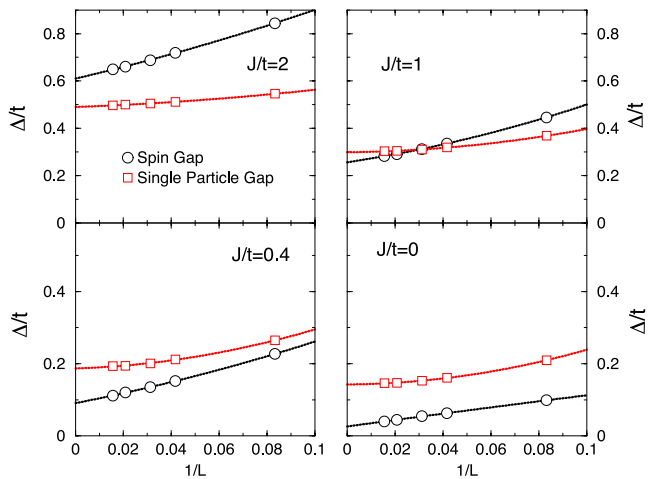}
  \caption{
   (Color online)
    DMRG results for the spin gap and single particle gap for $J/t=0, 0.4, 1, 2$,
    as a function of inverse system size $1/L$. 
    \label{fig:dmrggaps}}
\end{figure}
Next we address the question of the excitation gaps in the symmetry broken phase. The theoretical
picture predicts an insulating state with a finite gap to all excitations above the two-fold degenerate 
ground state. We calculate the single particle charge gap
and the spin gap defined in equations (\ref{eqn:1Pgap}) and (\ref{eqn:SpinGap}), respectively.
The calculated gaps are shown in Fig.~\ref{fig:dmrggaps}. The finite size gaps are extrapolated 
to $L=\infty$ with a simple quadratic fit. All gaps extrapolate to a finite value, in agreement
with the predictions. The charge gap more or less follows the increase of the
alternation of the kinetic energy shown in Fig.~\ref{fig:dmrgorderparameters},
i.e., the gap is roughly multiplied by a factor three going from $J/t=0$ to 2. The behavior
of the spin gap is mainly driven by the fact that it scales with $J/t$. Note that
even in the case $J/t=0$ the spin gaps seem to remain finite. 
It will be an interesting question to characterize the precise nature of the charge 
and spin excitations. This will be left for a future study.

\begin{figure}
  \includegraphics*[width=0.9\linewidth]{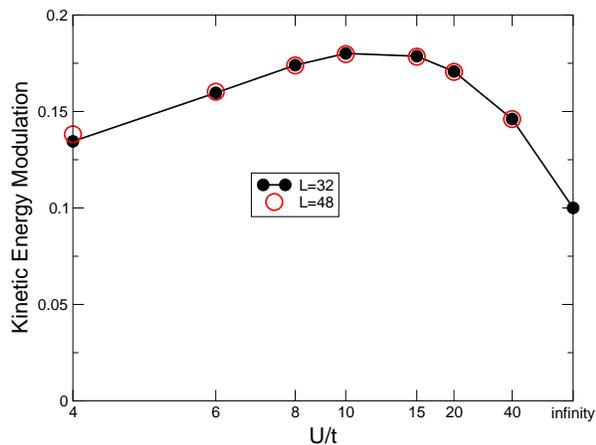}
  \caption{
    (Color online) 
    DMRG results for the kinetic energy alternation for Hubbard \kag strips at $n=2/3$ of 
    length $L=32$ and $L=48$.
    The modulation is non-monotonous as a function of $U/t$ and shows
    a maximum around $U/t\approx 10\sim15$.
    \label{fig:dmrghubbardorderparameter}}
\end{figure}
The weak coupling RG calculations in the following section are performed for Hubbard onsite interactions. Although we expect the
behavior of the $t{-}J$ model and the Hubbard model at large $U$ to be
similar, we have explicitly calculated the alternation of the kinetic 
energy for the Hubbard model as a function of $U/t$. The results displayed
in Fig.~\ref{fig:dmrghubbardorderparameter} show that this quantity has
a maximum around $U/t\approx 10\sim15$, and interpolates between the 
exponentially small order parameter at weak $U/t$ and the result for the
$t{-}J$ model at $J=0$, which corresponds to $U=\infty$. These results
therefore suggest that for the particular case of the \kag strip the weak-coupling
phase is adiabatically connected to the strong coupling limit.

\subsection{Weak-coupling discussion}
\label{weakcoupl}

We consider in this section the weak-coupling Hubbard model with the interaction
\begin{equation}
  \label{stripHintlocal}
  H_{\mathrm{Int}}=U\sum_{\br}
 c^{\dag}_{\br,\up}c^{\dag}_{\br,\down}c^{\phantom\dag}_{\br,\down}c^{\phantom\dag}_{\br,\up},
\end{equation}
on the 1D \kag strip. 
We map this  weak local Coulomb repulsion on the original \kag strip on an effective interaction for the underlying half filled two-leg ladder. 
We will show in the following that for the one-dimensional \kag strip, applying the RG and bosonization techniques that were developed for the half filled two-leg ladder, we find in fact a charge density wave (CDW) instability in the effective model for the two-leg ladder which corresponds to a BOW instability on the \kag strip.

\begin{figure}
  \includegraphics[width=\linewidth]{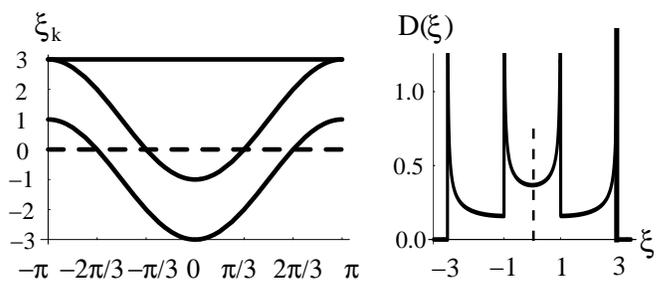}
  \caption{The \kag strip bands and the density of states per unit cell and spin. The energy is measured in units of $t$. \label{stripbands}} 
\end{figure}

The tight-binding bands of the \kag strip with $\mu=-t$ are shown in Fig.~\ref{stripbands}.
The dispersing bands are the same as the bands of a two-leg ladder. The flat band originates from states that are trapped
within one rhombus. The density of states has square-root singularities at
$\pm t,\pm 3t$ and a delta-peak at $3t$. The Fermi surface is given by the 4
points $\pm k_{\mathrm{F}1}$ and $\pm k_{\mathrm{F}2}$, where
$k_{\mathrm{F}1}=\pi/3$ and $k_{\mathrm{F}2}=2\pi/3$. There is a finite density
of states at the Fermi surface.
The \kag strip can be viewed as a \kag lattice tube, i.e., a \kag lattice with finite width and periodic
boundary conditions. 
 In order to see that the bands in Fig.~\ref{stripbands} are in fact a cut through the \kag dispersion shown in Fig.~\ref{kagomebands}, one has to shift one of the dispersing bands by $\pi$. 
This difference arises because our notation is chosen to emphasize the similarities of the \kag strip to
the two-leg ladder.

In contrast to the \kag and the pyrochlore lattice, the density of states at the Fermi surface is finite for the \kag strip and 
we therefore expect qualitative changes in this 1D system even for weak interactions.  
We perform a weak-coupling RG and bosonization
analysis for the \kag strip, and we show that the bond order wave instability is already present for arbitrary weak coupling. In this section we will only present the results of this analysis and refer to Appendix~\ref{kagappRG} for further details.

We derive an effective interaction for the two-leg ladder, that corresponds in weak coupling to the local Coulomb repulsion on the \kag strip (\ref{stripHintlocal}). 
In this derivation we can drop terms that involve the high energy states of the flat band and focus on the states  
in the dispersing bands. We denote the annihilation operator of these states by
$\g_{\bk,\sigma}=\g_{k,i,\s}$
where $k$ is the momentum along the strip and $i\in\{1,2\}$ is the band index.
If we rewrite the Hamiltonian $H_{\mathrm{Int}}$ in terms of these new operators we obtain the interaction.
\begin{equation}
  \label{Hintsquare}
  H_{\mathrm{Int}}\rightarrow\frac{U}{L}\sum_{\bk_1\dots\bk_4}^{\prime}g^{\phantom\dag}_{\bk_1\dots \bk_4}\,
  \g^{\dag}_{\bk_1,\up}\g^{\dag}_{\bk_2,\down}\g^{\phantom\dag}_{\bk_3,\down}\g^{\phantom\dag}_{\bk_4,\up},
\end{equation}
where the prime over the sum restricts the sum to momentum conserving $\bk$-values. 
 For weak interaction we can replace $\bk_l$ in $g_{\bk_1\dots\bk_4}$ by $(k_{\mathrm{F}i_l},i_l)$ and we obtain the simple expression 
\begin{equation}
g_{\bk_1\dots \bk_4}=e^{-i\frac{q}{2}}(\delta_{i_1i_2}\delta_{i_3i_4}+\delta_{i_1i_3}\delta_{i_2i_4}+\delta_{i_1i_4}\delta_{i_2i_3})/6,
\end{equation}
where $q=k_1+k_2-k_3-k_4$.

The effective interaction (\ref{Hintsquare}) can  now be expressed in terms of left and right moving currents and in this way we find the initial values for the RG equations  of the two-leg ladder.
The integration of the RG equations with these initial values
converges to an analytic solution that was identified by bosonization techniques as a charge density wave solution\cite{lin}. This means that
the operator
\begin{equation}
  \label{cdwO} 
O_{\mathrm{CDW}}=\frac{1}{L}\sum_{k,\sigma}\,\g^{\dag}_{k,1,\sigma}\g^{\phantom\dag}_{k+\pi,2,\sigma}+\g^{\dag
}_{k+\pi,2,\sigma}\g^{\phantom\dag}_{k,1,\sigma}
\end{equation}
acquires a finite value.
 The bond order wave order-parameter on the \kag strip is given by the expectation value of an operator $O_{\mathrm{BOW}}$.

The operators  $O_{\mathrm{CDW}}$ and $O_{\mathrm{BOW}}$ transform
identically under all symmetries of the system and, therefore,
they describe the same phase.

In addition, $O_{\mathrm{CDW}}$ is the effective operator on the two-leg ladder for $O_{\mathrm{BOW}}$, i.e., if one does the same substitutions as we did for deriving Eq.~(\ref{Hintsquare}) one sees that
$
O_{\mathrm{BOW}}\rightarrow  O_{\mathrm{CDW}}
$,
if one chooses the right prefactor in the definition of $O_{\mathrm{BOW}}$.

We have shown that the bond order wave instability that is expected to
occur at rather strong interactions according to the arguments of the
preceding sections, is in fact already present in weak coupling for
the one-dimensional \kag strip. Together with the DMRG results of the preceding section
we provide convincing evidence, that the BOW phase on the \kag strip is present both in the Hubbard model for all values of $U$ and in the \tJ model.  

\section{Conclusions}
\label{sec:summary}

In summary we have studied the occurrence of a bond order wave instability in several bisimplex lattices.
We provided evidence that this instability occurs quite generally in these lattices at the fractional filling of one electron per simplex (two electrons per ``up''-simplex), 
if the correlations -- i.e., antiferromagnetic nearest-neighbor exchange and/or nearest-neighbor repulsion -- are strong enough. 

In weak-coupling the physical properties of the system are dominated by the dimensionality of the lattice, by its fermiology and by the density of states at the Fermi energy.
We show that in the intermediate coupling regime, where the kinetic and the interaction energies are comparable, at the filling with two electrons per ``up''-simplex, the 
physical properties of these highly frustrated lattices are dominated by local states on the simplex. The bipartite and corner-sharing arrangement of the simplices 
allows the creation of isolated or only weakly interacting simplices with low energy by spontaneously breaking inversion or translational symmetry.
This knowledge provides a good starting point for series expansions or further CORE calculations.
  
The magnetic interaction and the chosen sign of the dispersion leads to a tendency to form nearest-neighbor singlets and nearest-neighbor repulsion leads to a
 tendency to avoid configurations with more than two electrons per simplex. If the underlying lattice is bipartite the system finds a way to satisfy both tendencies 
simultaneously by localizing singlets on every second simplex. This localization leads only to a partial loss of the kinetic energy, because the singlets can still 
delocalize within the simplex. It is the cooperation between the kinetic and the interaction energy which stabilizes the bond order wave state. 
Note, that the bond order wave instability does not lead to an inhomogeneous charge distribution on the lattice.

The bond order wave states, which we find on the different lattices, provide a natural generalization of the well-known valence bond solid states (e.g.\ dimerized phases,
plaquette phases) found in many frustrated spin models to situations away from half filling where a description in terms of spin variables only breaks down. The density
is still a rational fraction, but $n=2/3$ in the \kag  and \kag strip case while $n=1/2$ in the pyrochlore and the checkerboard case. Approximately these states are direct products
of singlets on triangles or tetrahedra, similar to the conventional picture of a dimerized phase. In contrast to the phases at $n=1$ however, the present instability involves a
cooperative effect of both magnetic exchange {\em and} kinetic energy.

An interesting task is to study the properties of a lightly doped bond order wave phase. It can be assumed that the bond order wave order parameter decreases 
rather quickly with doping. However, it is conceivable that away from the commensurate filling the bond order wave order parameter coexists with a small 
superconducting order parameter. This phase would at the same time break lattice symmetries and the $U(1)$ gauge symmetry and would be therefore 
similar to a supersolid. A closer investigation of these issues has however be left for further studies. 

In general, we conclude that the bond order wave instability occurs quite generally in bisimplex lattices and for physically reasonable models and 
interaction parameters. We show that doping frustrated spin models can lead to new phases. Our study may also be a step towards the understanding of the interplay of frustrated magnetic fluctuations and itinerant charge carriers, which could play a role for example in the unconventional heavy Fermion material LiV$_2$O$_4$~\cite{KondoLiV2O4} or in Na$_x$CoO$_2$.

\acknowledgments
We thank T.M.~Rice, K.~Wakabayashi, and D.~Poilblanc for stimulating discussions.
We acknowledge support by the Swiss National Fund and NCCR MaNEP.
Computations were performed on IBM Regatta machines of CSCS Manno and IDRIS Orsay.

\appendix
\section{RG analysis}

\subsection{\kag strip\label{kagappRG}}

The tight-binding Hamiltonian for the \kag strip with periodic boundary conditions is given by
\begin{eqnarray}
  \label{striptb}
  \hat{H}_0&=&-t\sum_{r=1}^L
\sum_{\nu=\pm1}\sum_{\s}\big[c^{\dag}_{r,\nu,\s}c^{\phantom\dag}_{r+\nu,\nu,\s}+\mbox{}\\
  && 
\mbox{}+c^{\dag}_{r,0,\s}(c^{\phantom\dag}_{r,\nu,\s}+c^{\phantom\dag}_{r+\nu,\nu,\s})+\hc\big]-\mu\hat{N}
\nonumber,
\end{eqnarray}
where $\s\in\{\up,\down\}$ are the spin indices and $\hat{N}$ is the number
operator.
The chemical potential, $\mu$, will be fixed to $-t=-1$ in the following. 
It is convenient to introduce Fourier transformed operators
\begin{eqnarray}
  \label{stripfto}
  c_{r,x}=\frac{1}{\sqrt{L}}\sum_k e^{ik(r-\frac{x}{2})}c_{k,x} \qquad
x\in\{\pm1,0\},
\end{eqnarray}
where the $k$-sum runs over the $L$ $k$-values in $[-\pi,\pi)$.
We can write this Hamiltonian in a diagonal form
\[\hat{H}_0=\sum_{k}\sum_{l=1}^3\sum_{\s}
\xi^{\phantom\dag}_{k,l}\,\g^{\dag}_{k,l,\s}\g^{\phantom\dag}_{k,l,\s}\label{stdiag},\]
and obtain the energies 
\begin{equation}
  \label{stenergies}
  \xi_{k,1}= 1-2\cos(k),\ \xi_{k,2}=-1-2\cos(k),\ \xi_{k,3}=3.
\end{equation}
and the operators 
\begin{equation}
  \label{dofc}
  \left[\begin{array}{c}\g_{k,1,\s}\\\g_{k,2,\s}\\\g_{k,3,\s}\end{array}
\right]=\frac{1}{\sqrt{2}\beta_k}
  \left[\begin{array}{ccc}
      \b_k&0&-\b_k\\
      \a_k&\sqrt{2}&\a_k\\
      1&-\sqrt{2}\a_k&1
    \end{array}\right]\!\!
  \left[\begin{array}{l}c_{k,-1}\\c_{k,0}\\c_{k,+1}\end{array} \right]
\end{equation}
with $\alpha_k=\sqrt{2}\cos(k/2)$ and $\beta_k=\sqrt{1+\a_k^2}$ and
$k\in[-\pi,\pi)$.

The local Coulomb interaction introduced in Eq.~(\ref{stripHintlocal}) can be written as
\begin{equation}
  \label{stripHint}
  H_{\mathrm{Int}}=\frac{U}{L}\sum^{\prime}_{k_1\dots k_4}
\sum_{x=-1}^1e^{-i\frac{xq}{2}}c^{\dag}_{k_1,x,\up}c^{\dag}_{k_2,x,\down}c^{\phantom\dag}_{k_3,
x,\down}c^{\phantom\dag}_{k_4,x,\up}.
\end{equation}
The sum over the momenta $k_1\dots k_4$ is restricted, such
that $q=k_1+k_2-k_3-k_4$ is a multiple of $2\pi$. Note, that the appearing phase
factor is important to determine the sign of the Umklapp scattering processes
correctly. 
We obtain the effective low-energy Hamiltonian (\ref{Hintsquare}) from
Eq.~(\ref{stripHint}) by doing the substitutions
\begin{eqnarray}
  \label{subs}
  c_{k,\pm,\s}&\rightarrow
  &\mp\frac{1}{\sqrt{2}}\g_{k,1,\s}+\frac{1}{\sqrt{6}}\g_{k,2,\s}\\ 
  c_{k,0,\s}&\rightarrow &\sqrt{\frac{2}{3}}\g_{k,2,\s}.\nonumber
\end{eqnarray}
These substitutions rules are obtained from Eq.~(\ref{dofc}) if we set $k=k_{\mathrm{F}1}$ in the first row and $k=k_{\mathrm{F}2}$ in the second row and drop the third row in the matrix of the transformation.

In this way we can map the weak-coupling Hubbard model on the \kag strip on an effective weak-coupling model on the two-leg ladder.
The problem of a weak-coupling two-leg ladder has been extensively studied by
renormalization-group and bosonization
techniques\cite{lin,fjaerestad,tsuchiizu}.
We will adopt here the notation of Lin \et\ in  Ref.~\onlinecite{lin}.
A general weak interaction can be conveniently expressed in terms of left and
right moving currents.
Dropping purely chiral terms the momentum-conserving four fermion interactions
can be written as    
\begin{eqnarray}
\mathcal{H}^{(1)}&=&b^{\rho}_{ij}J_{Rij}J_{Lij}-b^{\s}_{ij}\mathbf{J}_{Rij}
\cdot\mathbf{J}_{Lij}\\
&+&
f^{\rho}_{ij}J_{Rii}J_{Ljj}-f^{\s}_{ij}\mathbf{J}_{Rii}\cdot\mathbf{J}_{Ljj}.
\nonumber
\end{eqnarray} 
To avoid double counting we set $f_{ii}=0$. Furthermore, the symmetry relations
$f_{12}=f_{21}$ (parity), 
$b_{12}=b_{21}$ (hermicity), and $b_{11}=b_{22}$ (only at half filling) hold. We have
therefore six independent coefficients.
For our interaction we find the values
\[4b_{11}^{\rho}=b_{11}^{\s}=U, \quad4b_{12}^{\rho}=b_{12}^{\s}=\frac{U}{3}, \quad
4f_{12}^{\rho}=f_{12}^{\s}=\frac{U}{3}. \]
In addition we have Umklapp terms given by
\[
\mathcal{H}^{(2)}=u_{ij}^{\rho}I^{\dag}_{Rij}I_{L\bar{i}\bar{j}}-u_{ij}^{\s}
\mathbf{I}^{\dag}_{Rij}\cdot\mathbf{I}_{L\bar{i}\bar{j}}+\hc\]with
$u_{ii}^{\s}=0$, $u_{11}^{\rho}=u_{22}^{\s}$ and $u_{12}^{\rho}=u_{21}^{\rho}$
and $u_{12}^{\s}=u_{21}^{\s}$.
Here we  have the values
\[u_{12}^{\s}=0,\quad u_{12}^{\rho}=-\frac{U}{12},\quad
u_{11}^{\rho}=-\frac{U}{24}.\]
Integrating the RG equations with these initial values shows that the solution
converges to the analytic solution of the RG equation where all coupling
constants except for $b^{\s}_{11}$ and $b^{\r}_{11}$ diverge with fixed ratios
given by
\begin{eqnarray}
\label{cdwfp}
 f^{\r}_{12}&=&-\frac{1}{4}f^{\s}_{12}=b^{\r}_{12}=-\frac{1}{4}b^{\s}_{12}\\
&=&\frac{1}{2}u^{\s}_{12}
=-2u^{\r}_{12}=-2u^{\r}_{11}=g>0.\nonumber
\end{eqnarray}
This solution was identified by bosonization techniques as a charge density wave
solution (CDW) solution.

\end{document}